\documentclass{elsart}
\usepackage{graphicx}
\usepackage{epsfig}

\def \vec #1{\mbox{{\boldmath $#1$}}}
\def \cos {\rm cos}

\def \costh {|\cos\:\theta^{*}|}
\def \sin {\rm sin}
\def \GeV {~{\rm GeV}}

\def \gg {\gamma\gamma}
\def \ggqqqq {\gamma\gamma\to q\bar{q}q\bar{q}}
\def \mumu {\mu^{+}\mu^{-}}
\def \tautau {\tau^{+}\tau^{-}}
\def \pipi {\pi^{+}\pi^{-}}

\def \KK {K^{+}K^{-}}

\def \MM {M^{+}M^{-}}
\def \ee {e^{+}e^{-}}
\def \qq {q\overline{q}}

\def \sigmaK {\sigma(\gg\rightarrow K^{+}K^{-})}
\def \sigmapi {\sigma(\gg\rightarrow \pi^{+}\pi^{-})}
\def \ratioKpi {\sigmaK/\sigmapi}
\def \ptbal {|\vec{p}_{t}^{+} + \vec{p}_{t}^{-}|}
\def \ggqq {\gg\rightarrow \qq}
\def \ggKK {\gg\rightarrow \KK}
\def \ggMM {\gg\rightarrow \MM}
\def \ggpipi {\gg\rightarrow \pipi}
\def \ggmumu {\gg\rightarrow \mumu}

\def \eetautau {\ee\rightarrow\tautau}
\def \cosre {\costh < 0.6}
\def \highr {3.0\,{\rm GeV} < W < 4.1\,{\rm GeV}}

\def \massr {2.4\,{\rm GeV} < W < 4.1\,{\rm GeV}}
\def \Ggg {\Gamma_{\gamma\gamma}}
\def \Lgg {{\it L}_{\gg}}
\def \Ldt {{\int\!\!{\mathcal{L}}}dt}
\def \dsdW {\frac{d\sigma}{dW}}
\def \prob {\mathcal{R}}
\def \sinnf {\sin^{-4}\:\theta^{*}}
\def \sinf {\sin^{4}\:\theta^{*}}
\newcommand{\W}{{\it W}}

\newcommand{\ValsRKK} {0.68 \pm 0.01 \pm 0.05~\GeV^2 }
\newcommand{\ValsRPI} {0.71 \pm 0.01 \pm 0.05~\GeV^2 } 

\newcommand{\ValIntLumi}{87.7\,{\rm fb}^{-1}}
\newcommand{\ValPIGggBrA}{15.1 \pm 2.1 \pm 2.3} 
\newcommand{\ValPIGggBrB}{0.76 \pm 0.14 \pm 0.11}
\newcommand{\ValKKGggBrA}{14.3 \pm 1.6 \pm 2.3}
\newcommand{\ValKKGggBrB}{0.44 \pm 0.11 \pm 0.07}

\newcommand{\ValCombGggA}{ 2.62 \pm 0.23(stat.) \pm 0.31(syst.) \pm 0.24(\mathcal{B})~{\rm keV} }

\newcommand{\ValCombGggB}{ 0.44 \pm 0.07(stat.) \pm 0.05(syst.) \pm 0.05(\mathcal{B})~{\rm keV} }

\newcommand{\ValSigmaRatio} {0.89 \pm 0.04(stat.) \pm 0.15(syst.) }

\newcommand{\ValKKNCHIA}{153}
\newcommand{\ValKKNCHIB}{33}
\newcommand{\ValPINCHIA}{129}
\newcommand{\ValPINCHIB}{54}
\newcommand{\ValKKNCHIAwithE}{153 \pm 17}
\newcommand{\ValKKNCHIBwithE}{33 \pm 8}
\newcommand{\ValPINCHIAwithE}{129 \pm 18}
\newcommand{\ValPINCHIBwithE}{54 \pm 10}
\newcommand{\ValPISIGA}{6.2\sigma}
\newcommand{\ValPISIGB}{4.8\sigma}
\newcommand{\ValKKSIGA}{8.2\sigma}
\newcommand{\ValKKSIGB}{3.7\sigma}

\newcommand{\ValPIPowers}{-7.9 \pm 0.4 \pm 1.5}
\newcommand{\ValKKPowers}{-7.3 \pm 0.3 \pm 1.5}

\begin{document}
\begin{frontmatter}

% <<< for preprint >>>
\vspace*{-3\baselineskip}
\begin{flushleft}
 \resizebox{!}{3cm}{\includegraphics{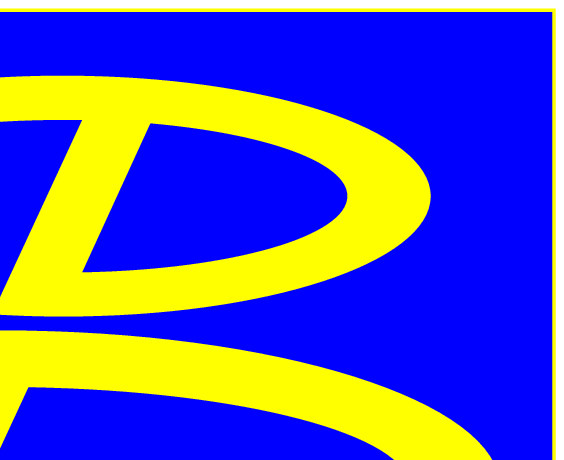}}
\end{flushleft}
\vspace*{-3cm}
\begin{flushright}
 Belle Preprint 2004-39\\
 KEK Preprint 2004-80 
\end{flushright}
\vspace*{2cm}

\title{
Measurement of the $\ggpipi$ and $\ggKK$
processes \\at energies of 2.4--4.1\,GeV
}
%%% Paper:    g g -> K K, g g -> pi pi
%%% Journal:  Physics Letters B
%%% Contacts: H. Nakazawa (nkzw@bmail.kek.jp)
%%% Non-responding authors or those who said NO are commented out.
%%% ====================================================================
%%% Click the RELOAD button on your web browser to see the updated file.
%%% ====================================================================
%%% Use \input{author} to insert this material into your latex file.
\collab{Belle Collaboration}
  \author[KEK]{H.~Nakazawa}, % KEK
  \author[KEK]{S.~Uehara}, % KEK
  \author[KEK]{K.~Abe}, % KEK
  \author[TohokuGakuin]{K.~Abe}, % TohokuGakuin
% \author[TIT]{N.~Abe}, % TIT
% \author[KEK]{I.~Adachi}, % KEK
  \author[Tokyo]{H.~Aihara}, % Tokyo
  \author[Nagoya]{M.~Akatsu}, % Nagoya
  \author[Tsukuba]{Y.~Asano}, % Tsukuba
% \author[Toyama]{T.~Aso}, % Toyama
  \author[BINP]{V.~Aulchenko}, % BINP
  \author[ITEP]{T.~Aushev}, % ITEP
% \author[Tata]{T.~Aziz}, % Tata
  \author[Cincinnati]{S.~Bahinipati}, % Cincinnati
% ...
% Ljubljana
% \author[Lausanne]{D.~Z\"urcher}, % Lausanne

% ...
\address[BINP]{Budker Institute of Nuclear Physics, Novosibirsk, Russia}
% ...
\address[Cincinnati]{University of Cincinnati, Cincinnati, OH, USA}
% ...
\address[KEK]{High Energy Accelerator Research Organization (KEK), Tsukuba, Japan}
% ...
\address[ITEP]{Institute for Theoretical and Experimental Physics, Moscow, Russia}
% ...
\address[Nagoya]{Nagoya University, Nagoya, Japan}
% ...
\address[TohokuGakuin]{Tohoku Gakuin University, Tagajo, Japan}
% ...
\address[Tokyo]{Department of Physics, University of Tokyo, Tokyo, Japan}
% ...
\address[Tsukuba]{University of Tsukuba, Tsukuba, Japan}
% ...
\thanks[NovaGorica]{on leave from Nova Gorica Polytechnic, Nova Gorica, Slovenia}

\newpage
\begin{abstract}
\quad We have measured 
$\pipi$ and $\KK$
production in two-photon collisions using $\ValIntLumi$ of data
collected with the Belle detector
at the asymmetric energy $\ee$ collider KEKB.
The cross sections are measured to high precision
in the two-photon center-of-mass energy ($W$) range
between $\massr$ and
angular region 
$\cosre$.
The cross section ratio $\ratioKpi$
is measured to be 
$\ValSigmaRatio$ 
in the range of $\highr$, where the ratio 
is energy independent.
We observe a
$\sinnf$ behavior of the cross section
in the same $W$ range.
Production of $\chi_{c0}$ and $\chi_{c2}$ mesons is observed in both
$\ggpipi$ and $\ggKK$ modes.
\end{abstract}

\begin{keyword}
two-photon collisions \sep mesons \sep QCD \sep charmonium
\PACS 12.38Qk \sep 13.25.Gv \sep 13.66.Bc \sep 13.85.Lg
\end{keyword}
\end{frontmatter}

\section{Introduction}
Exclusive processes with hadronic final states test
various model calculations motivated 
by perturbative or non-perturbative QCD.
Two-photon production of exclusive hadronic final states is
particularly attractive due to the absence of strong interactions 
in the initial state and the possibility of calculating $\gg\to\qq$ amplitudes.
The perturbative QCD calculation by Brodsky and Lepage (BL)~\cite{Lepage}
is based on factorization of the amplitude into a hard scattering amplitude
for $\ggqqqq$ and a single-meson distribution amplitude.   
Their prediction gives the dependence on the center-of-mass (c.m.)\
energy $W(\equiv\!\sqrt{s})$ and scattering angle $\theta^*$ for $\ggMM$ processes
\begin{eqnarray}
  \hspace{2cm}
  \frac{d\sigma}{d\costh}(\ggMM)
  \approx
  \frac{16\pi\alpha^2}{s}
  \frac{|{\it F}_M(s)|^2}{\sinf},
  \label{eq:brod}
\end{eqnarray}
where $M$ represents a meson and $F_M$ denotes its 
electromagnetic form factor. 
%Vogt, based on the perturbative approach,
%showed~\cite{vogt} that the hard contribution is well below the
%experimental cross section obtained by CLEO~\cite{cleo}
%and claimed a need for sizeable soft contributions,
%in the finite energy region.
Vogt~\cite{vogt}, based on the perturbative approach, claimed
a need for soft contributions,
as his result for the hard contribution was well below the
experimental cross section obtained by CLEO~\cite{cleo}.

Diehl, Kroll and Vogt (DKV)
proposed~\cite{handbag} the soft handbag contribution 
to two-photon annihilation into pion or kaon pairs 
at large energy and
momentum transfers, in which 
the amplitude is expressed 
by a hard $\ggqq$ subprocess 
and a form factor describing 
the soft transition from $\qq$ to
the meson pair.
DKV, as well as BL, predict 
the $\sinnf$ dependence of
the angular differential cross section,
which is an important test of these approaches.
It is interesting to investigate experimentally an energy scale
where those theoretical predictions become valid.
The recent measurements of $\ggpipi$ and $\KK$
performed by ALEPH~\cite{aleph}
are consistent, within their errors, with the BL's prediction of 
the energy dependence, but not the normalization.  
However, their dataset
is not sufficient to conclusively test the $W$ and $\sinnf$ dependences.

In this report, we measure 
$\ggpipi$ and $\ggKK$ processes
with high precision, and make quantitative comparisons with
QCD predictions.
This analysis is based on an $\ValIntLumi$ data sample 
collected at or near the $\Upsilon(4S)$ resonance energy,
accumulated with
the Belle detector~\cite{belle} located at KEKB~\cite{kekb}.

\section{KEKB accelerator and Belle detector}

KEKB is a colliding beam accelerator of 
8$\GeV$  electrons and 3.5$\GeV$  positrons designed to produce
copious $B\overline{B}$ meson pairs
to observe {\it CP} violation.

The Belle detector, with a $1.5\,\rm T$ solenoidal magnetic field, surrounds
the interaction point and subtends the polar angle range $17^{\circ} < \theta_{\rm lab}
< 150^{\circ}$, measured from 
the $z$ axis, which is aligned opposite the positron beam.
It is described in
detail in Ref.~\cite{belle}.  Briefly, charged track momenta and their decay
points are measured by the central drift chamber (CDC) and silicon vertex
detector (SVD).  The hadron identity of these charged tracks is determined
using information from the time-of-flight counters (TOF), the aerogel
threshold \v Cerenkov counters (ACC), and the specific ionization in the CDC.
Hadron/electron discrimination is performed using the above information as
well as the energy deposition and shower profile in the segmented
CsI(Tl)-crystal electromagnetic calorimeter (ECL).  Hadron/muon discrimination
is achieved using information from the neutral-kaon and muon detector (KLM),
which consists of glass resistive plate counters embedded in the solenoid's
iron flux return.

\section{Event selection}\label{select}
The signal events are collected predominantly by a trigger that requires two
charged tracks penetrating through the CDC and TOF, with an opening angle in
the $r\varphi$ plane (perpendicular to the $z$ axis) of at least $135^{\circ}$.

We select signal candidates according to the following criteria.  There must
be exactly two oppositely-charged reconstructed tracks satisfying the
following conditions:  $-0.47 \le \cos\;\theta_{\rm lab} \le 0.82$ for the polar
angle $\theta_{\rm lab}$ of each track; $p_t > 0.8$ GeV/$c$ for the momentum component
in the $r\varphi$ plane of each track; $dr \le 1$ cm and $|dz| < 2$ cm for the
origin of each track relative to the nominal $e^+e^-$ collision point; and
$|dz_1 - dz_2| \le 1$ cm for the two tracks' origin difference along the
$z$ axis, where the origin is defined by the closest approach of the track
to the nominal collision point in the $r\varphi$ plane.  
The event is vetoed if it contains any other reconstructed
charged track with transverse momentum above 0.1 GeV/$c$.

Cosmic rays are suppressed by demanding that the opening angle $\alpha$
   between the two tracks satisfy $\cos\alpha \geq -0.997$.  The signal is
   enriched relative to other backgrounds by requiring that the scalar sum of the
   momentum of the two tracks be below 6 GeV/$c$, the total energy deposited
   in the ECL be below 6 GeV, the magnitude of the net transverse momentum
   of the above-selected two charged tracks in the $e^+e^-$ c.m.\ frame 
   %calculated with mass-less transition 
   be below 0.2 GeV/$c$ 
   (this condition is tightened in the next
   section), the invariant mass of these two tracks be
   below 4.5 GeV/$c^2$, and that the squared missing mass of the event be above
   2 GeV$^2$/$c^4$. 
   Here, the two tracks are assumed to be massless particles.
   The latter two requirements eliminate radiative Bhabha and
   initial state radiation events.  
   The remaining events
   consist of two-photon production of $e^+e^-$, $\mu^+\mu^-$, $\pi^+\pi^-$,
   $K^+K^-$, and $p\bar{p}$ final states as well as unvetoed $\eetautau$ 
   events
   according to a Monte Carlo (MC) study~\cite{tautau}.

\begin{figure}[t]
  \begin{tabular}{cc}
    \begin{minipage}{0.5\hsize}
      \begin{center}
	\includegraphics[scale=0.32]{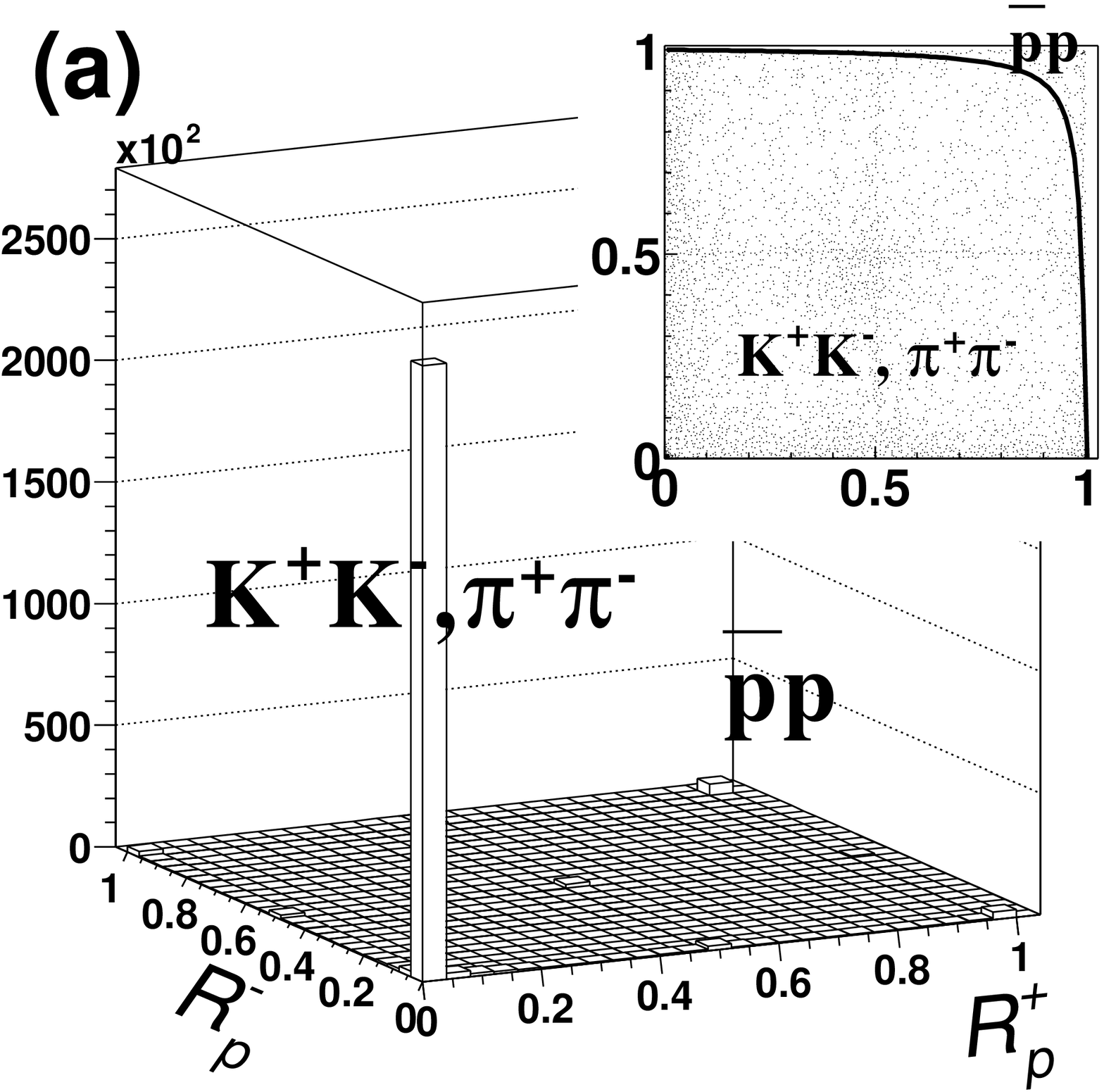}
      \end{center}
    \end{minipage}
    \begin{minipage}{0.5\hsize}
      \begin{center}
	\includegraphics[scale=0.32]{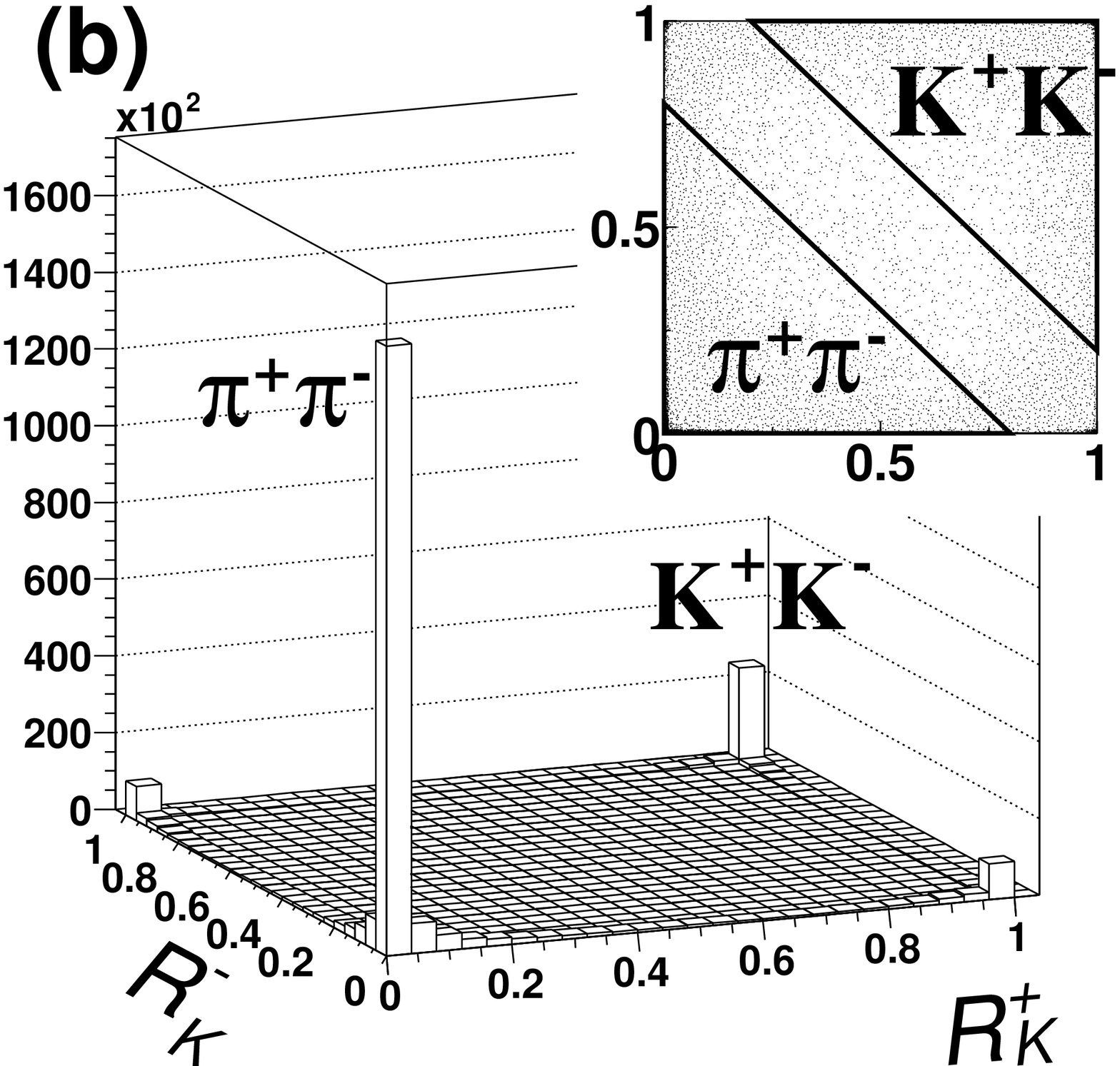}
      \end{center}
    \end{minipage}
  \end{tabular}
  \caption{
    Two-dimensional plots of 
    likelihood ratios for hadron identification:
    (a) $\prob$$_p$ and
    (b) $\prob$$_{K}$.
    The cut boundaries are shown in the top-view insets.
  }
  \label{fg_sel}
\end{figure}

The predicted versus measured range and transverse deviation of hits in the
   KLM are used to construct a normalized likelihood $\mathcal{R}_\mu$ that a
   track extrapolated from the CDC is a muon rather than a pion or kaon.
   Here, we classify an event as arising from $\gamma\gamma \to \mu^+\mu^-$ if
   either track has $\mathcal{R}_\mu > 0.66$.  Similarly, the TOF, ACC, CDC, and
   ECL information is used to construct a normalized likelihood $\mathcal{R}_e$
   that a reconstructed track is an electron rather than a hadron.  We classify
   an event as arising from $\gamma\gamma \to e^+e^-$ if either track has
   $\mathcal{R}_e > 0.66$.  
   For above two separations 93\% of signal events survive for both modes
   according to the MC study described later.
   The TOF, ACC and CDC information is used to construct
   another normalized likelihood $\mathcal{R}_p$ that a reconstructed track is a
   proton rather than a kaon, with a high value corresponding to a proton-like
   track.  The scatterplot of this quantity for the negative track {\it vs.} 
   that
   for the positive track in each event is shown in Fig.~\ref{fg_sel}(a). 
   Note that the peak near the origin contains both $\KK$ and $\pipi$ 
   candidates.
   Events above the hyperbolic curve 
   $(\mathcal{R}_p^- - 1.01)(\mathcal{R}_p^+ - 1.01) = 0.0101$, 
   shown in the inset of Fig.~\ref{fg_sel}(a), are deemed to arise from
   $\gamma\gamma \to p\bar{p}$.

%figsel

After removing events that appear to arise from two-photon production of
   $\mu^+\mu^-$, $e^+e^-$, and $p\bar{p}$ according to the above criteria, the
   remaining sample consists of two-photon production of $K^+K^-$, $\pi^+\pi^-$,
   and residual $\mu^+\mu^-$, as well as $\eetautau$ production
   where each $\tau$ lepton decays to a single pion or muon.  Information from
   the TOF, ACC, and CDC is used to form a normalized likelihood ratio
   $\mathcal{R}_K$ that a reconstructed track is a kaon rather than a pion (or
   muon), with a high value corresponding to a kaon-like track.  The scatter
   plot of this quantity for the negative track {\it vs.} that for the positive track
   is shown in Fig.~\ref{fg_sel}(b).  
   Note that the peak near the origin contains
   $\pipi$, $\tautau$, and residual $\mu^+\mu^-$ events.  
   Events above the diagonal
   line $\mathcal{R}_K^- + \mathcal{R}_K^+ = 1.2$, shown in the inset of Fig.~\ref{fg_sel}(b),
   are classified as $\KK$ candidates, while events below the line
   $\mathcal{R}_K^- + \mathcal{R}_K^+ = 0.8$ are classified as $\pipi$ candidates
   (including $\tautau$ and residual $\mumu$ backgrounds).  
   Events in the diagonal band between these two lines are discarded.

The $\pipi$ sample is somewhat contaminated by non-exclusive two-photon
   background $\ggpipi X$ as well as the $\eetautau$
   process, in roughly equal proportion.  
   We note that these
   backgrounds appear at high values of the 
   magnitude of the net transverse momentum $\ptbal$ in the $e^+e^-$ c.m.\ frame,
   and are often accompanied by photons from the prompt decay of a
   neutral pion in the final state.  Therefore, we reject events in the
   $\pipi$ sample that contain a photon with energy above 400 MeV
   ($E_\gamma$-veto).  The
   distributions of $\ptbal$ for the $\pipi$ candidates
   before and after application of this veto are shown 
   as the histograms in Fig.~\ref{fg:ptb}(a).

     The yields of the $\pi^+\pi^-$ and $K^+K^-$ events are 
     expressed as functions of three variables: $W$
     derived from the invariant mass of the two mesons, 
     $\costh$ and $\ptbal$.
     Eighty-five
    20 MeV wide bins in $W$ times
    six bins in the cosine of the $\gg$ c.m.\ scattering angle
    $\theta^*$ times twenty bins in net transverse momentum are used 
    in the ranges
    $\massr$, $\cosre$, and $\ptbal < 0.2\,{\rm GeV}/c$.
 \section{Background rejection}

The spectrum of the residual $\ggmumu$ background within
   the $\pipi$ sample is obtained from a MC simulation, based
   on a full $\mathcal{O}(\alpha^4)$ QED calculation~\cite{aafhb}, with a data
   sample corresponding to an integrated luminosity of $174.2\ \rm fb^{-1}$
   that is processed by the full detector simulation program and then subjected
   to trigger simulation and the above event selection criteria.  
   After calibration of the muon identification efficiency to match that
   in the data using identified $\ggmumu$ events, 
   the residual $\mumu$ background is scaled 
   by the integrated luminosity ratio and then subtracted.

\begin{figure}[th]
  \begin{center}
    \includegraphics[scale=0.65]{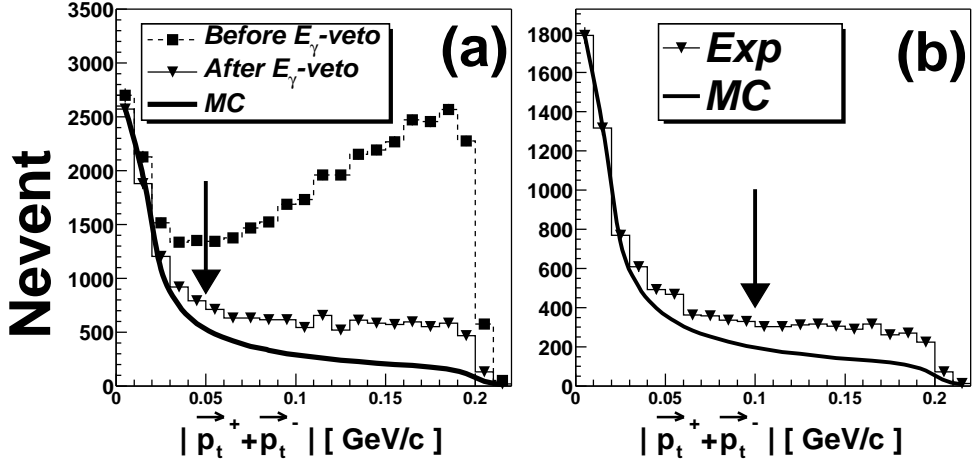}
    \caption{
      $|\vec{\textbf{p}}_{t}^{+} + \vec{\textbf{p}}_{t}^{-}|$ distribution 
      for $\pipi$(a) and $\KK$(b) candidates.
      The dashed and solid histograms in $\pipi$ indicate 
      the distribution of events
      before and after $E_\gamma$-veto
      (which is not applied to the $\KK$ candidates),
      respectively.
      The arrows indicate the upper boundaries of 
      $|\vec{\textbf{p}}_{t}^{+} + \vec{\textbf{p}}_{t}^{-}|$ for the signal.
      The residual muon background has been subtracted from the $\pipi$ distribution.
      The curves show the signal MC distribution
      which is normalized to the signal candidates at the leftmost bin.
    }
    \label{fg:ptb}
  \end{center}
\end{figure}

\begin{figure}[b]
  \begin{center}
    \includegraphics[scale=0.5]{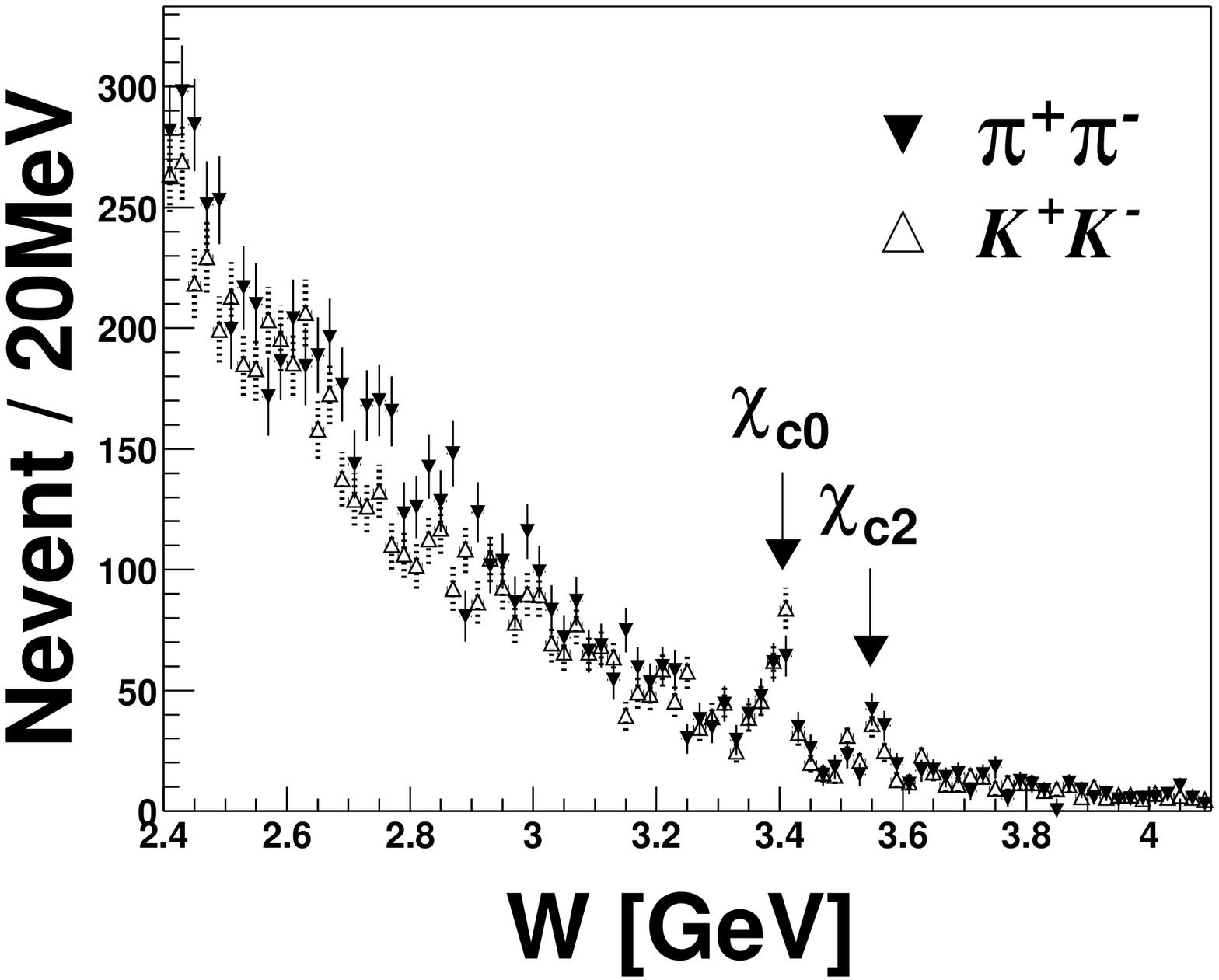}
    \caption{
      Number of events ($\cosre$) 
      obtained for the
      $\ggpipi$(solid) and
      $\ggKK$(dashed) samples
      after the background subtraction.
    }
    \label{fg:KKPIcrtw}
  \end{center}
\end{figure}

The excess in the $E_{\gamma}$-vetoed histogram of Fig.~\ref{fg:ptb}(a) 
   above the smooth curve from the signal MC---described in more detail below---is 
   attributed to
   non-exclusive $\gamma\gamma \to \pi^+\pi^- X$ events that are not rejected
   by the $E_\gamma$-veto; most of the $\eetautau$ events are rejected by this
   veto.  
   A similar excess appears in Fig.~\ref{fg:ptb}(b) for the
   $\ggKK$ process.  Assuming that this remaining background
   is proportional to net transverse momentum, we determine the slope using
   the difference between data and MC in the range $0 < \ptbal < 0.17$ GeV/$c$ 
   for each 200 MeV wide $W$ bin, then
   smooth the so-determined slopes by fitting them to a cubic polynomial in $W$.
   We verify that there is no dependence on the scattering angle $\theta^*$.
    Using the smoothed slope,
   the estimated non-exclusive background is subtracted from each bin.
Finally, we restrict our signal region to net transverse momentum below
   0.05 (0.10) GeV/$c$ for $\ggpipi$ ($\ggKK$).  
   The background-subtracted yields, integrated over net transverse
   momentum and scattering angle, 
   are shown as a function of $W$ in Fig.~\ref{fg:KKPIcrtw}.

   %figptb

The signatures of the
$\chi_{c0}(1P)$ and $\chi_{c2}(1P)$ resonances are observed
in both $\pipi$ and $\KK$ channels.
   By
   fitting each $W$ distribution outside the range 3.3--3.7 GeV
   to a cubic polynomial, we see an excess of $\ValPINCHIA$ ($\ValKKNCHIA$) events in the
   $\pipi$ ($\KK$) channel in the $\chi_{c0}$ range of 3.34--3.44 GeV
   and a corresponding excess of $\ValPINCHIB$ ($\ValKKNCHIB$) events in the 
   $\chi_{c2}$ range of
   3.54--3.58 GeV. We obtain consistent results from a fit of each
   distribution to a cubic polynomial plus a Breit-Wigner ($\chi_{c0}$)
   or a Gaussian ($\chi_{c2}$) peak.  Assuming a
   flat ($\sinf$) shape for 
   the $\chi_{c0}$ ($\chi_{c2}$) resonance~\cite{chi_c2},
   we subtract the above excesses bin by bin from each angular distribution
   in the above $W$ ranges.
   The $\chi_{c0}$ statistical significance is $\ValPISIGA$ 
   ($\ValKKSIGA$) in the $\pipi$ ($\KK$) channel, where $\sigma$ is
   the standard deviation.  The $\chi_{c2}$ statistical significance is
   $\ValPISIGB$ ($\ValKKSIGB$) in the $\pipi$ ($\KK$) channel.  
   The significances are taken from the square root of the difference
   of the goodness-of-fit values from the two fits where the
   peak term in the above fit function is included or excluded.
   %They are obtained by fitting the $W$-distribution
   %with and without a peak term of the function and taking the 
   %square root
   %of the difference of the goodness of fit values.
   %The Breit-Wigner (Gaussian) function and a cubic polynomial are used to 
   %describe the $\chi_{c0} (\chi_{c2})$ and the continuum, respectively.

\section{Derivation of the cross section}

 The differential cross section for a two-photon process 
 to a two-body final state
 arising from an electron-positron collision is given by
\begin{eqnarray}
  \hspace{-0.6cm}
 \frac{d\sigma}{d\costh}(W,\costh;\gg\to {\it X})
 =
 \frac
 {
 \Delta N(\W,\costh;
 {\it e^+}{\it e^-}\to{\it e^+}{\it e^- X})
 }
 {
 \Lgg(\W)\Delta\W
 \Delta \costh \epsilon(\W,\costh)\Ldt
 }
\end{eqnarray}
 where $N$ and $\epsilon$ denote 
 %the two-photon c.m.\ energy
 %determined by the invariant mass of the final-state system,
 the number of the signal events
 and a product of
 detection and trigger efficiencies, respectively;
 $\Ldt$ is the integrated
 luminosity, and $L_{\gg}$ is the luminosity function,
 defined as 
 $
 L_{\gg}(W)=
 {\dsdW}
 (W;\ee\!\to\!\ee X)/\sigma(W;\gg\!\to\! X)
 $.
 %%%

The efficiencies $\epsilon(W,\costh)$ for
   $\ggpipi$ and $\ggKK$ are obtained from a full Monte Carlo simulation~\cite{geant},
   using the TREPS~\cite{TREPS} program for the event generation as well as the
   luminosity function determination.  
   The trigger efficiency is determined from the trigger simulator.
   %assuming that it depends only on
   %the averaged transverse momentum of the two tracks.
   The typical value of the trigger efficiency is $\sim 93\%$
   for events in the acceptance.

\begin{center}
\begin{figure*}
 \includegraphics[scale=0.65]{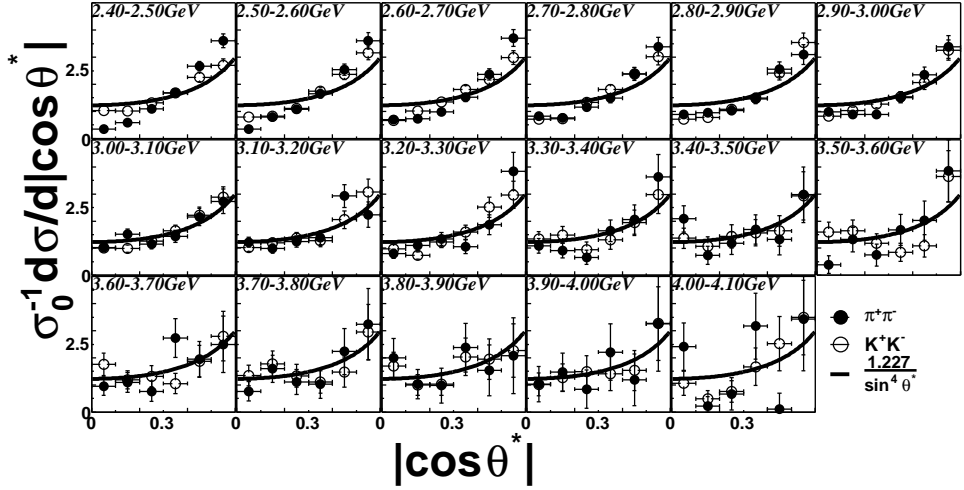}
 \caption{
 Angular dependence of the cross section,
 $\sigma_{0}^{-1}d\sigma/d\costh$,
for the $\pipi$(closed circles) 
and $\KK$(open circles) processes.
The curves are $1.227\times\sinnf$.
The errors are statistical only.
 }
 \label{fg:sin4}
\end{figure*}
\end{center}

\begin{center}
\begin{figure}[h]
  \epsfig{file=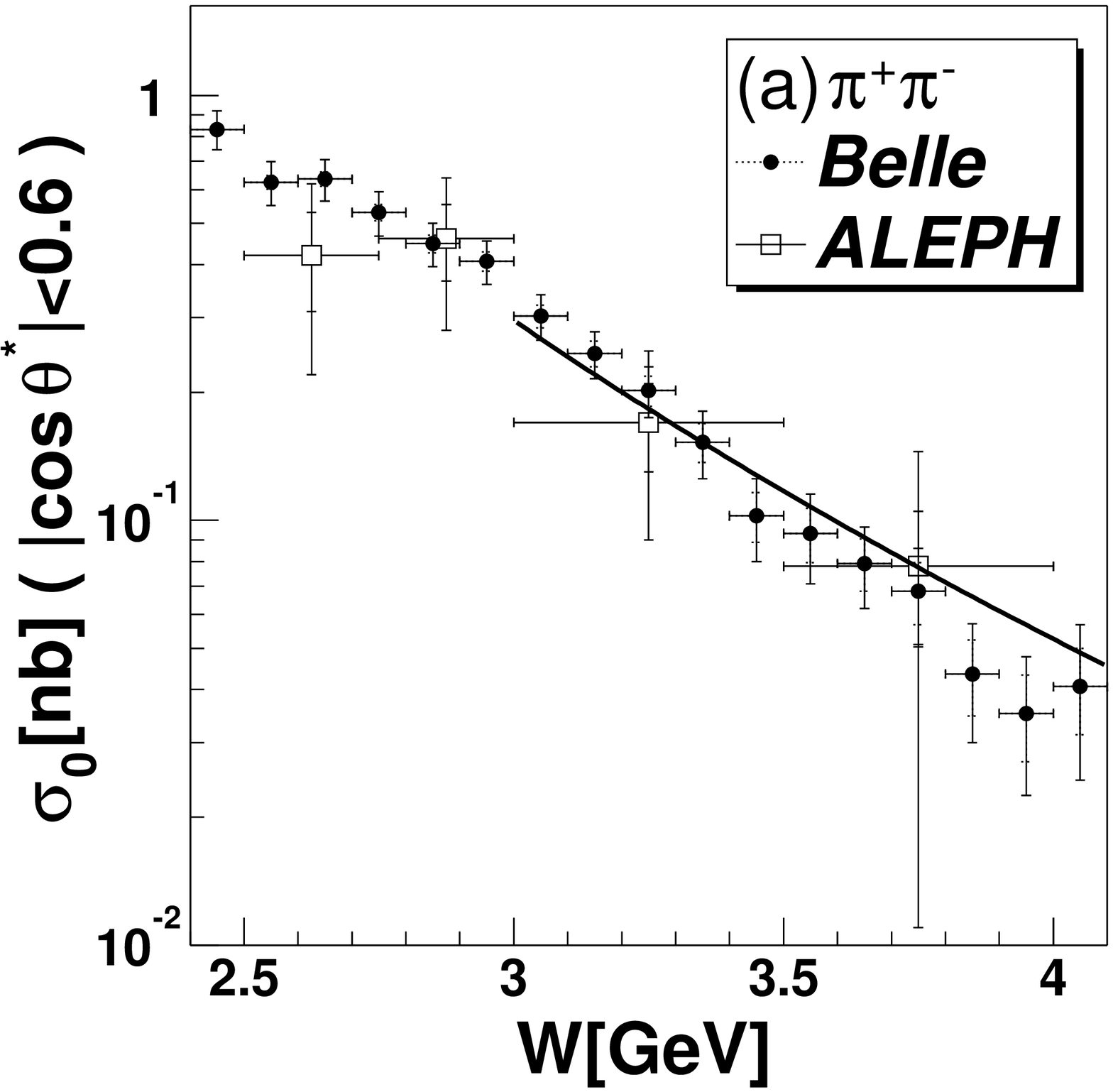,width=7cm}
  \epsfig{file=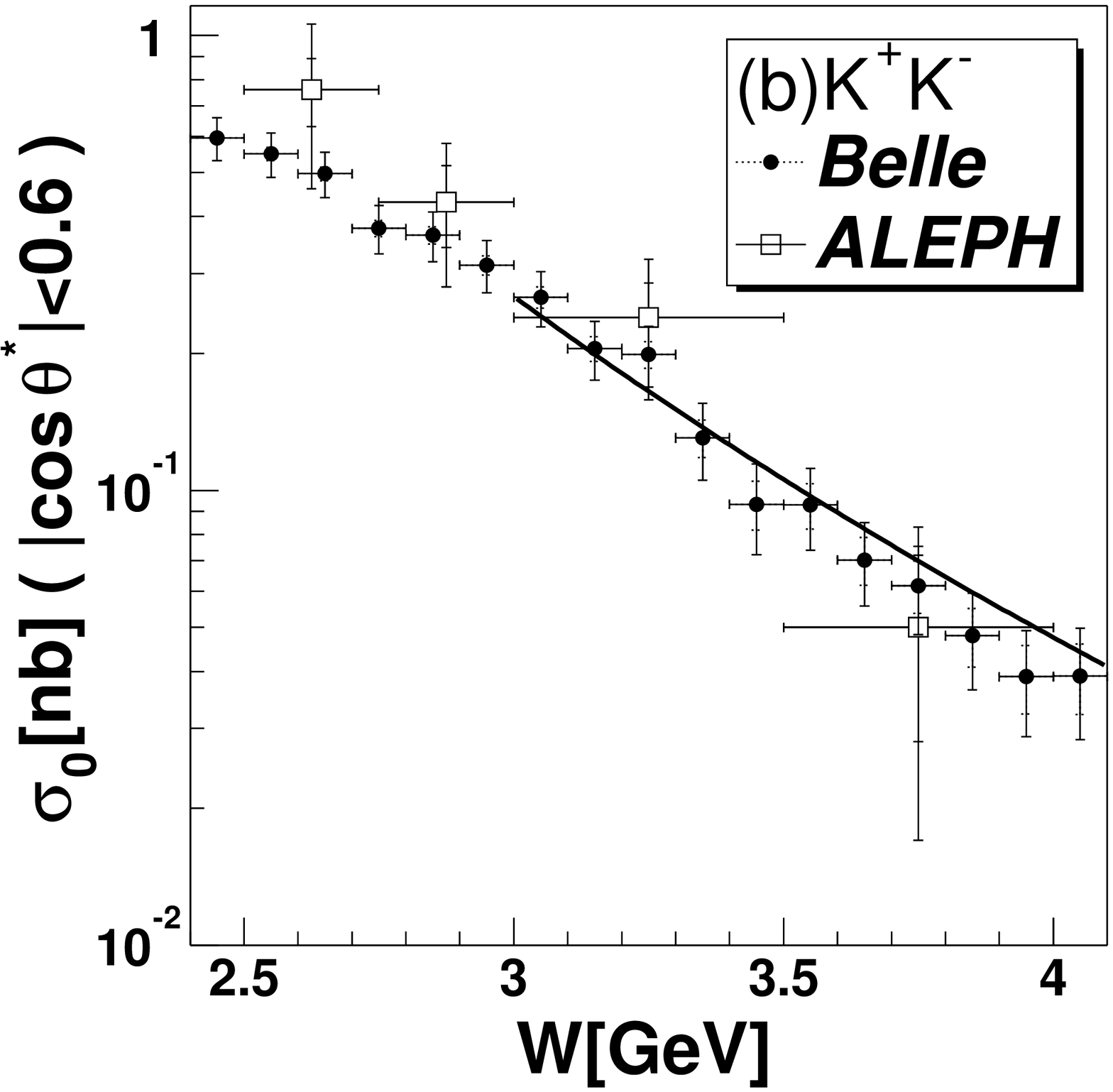,width=7cm}
  \epsfig{file=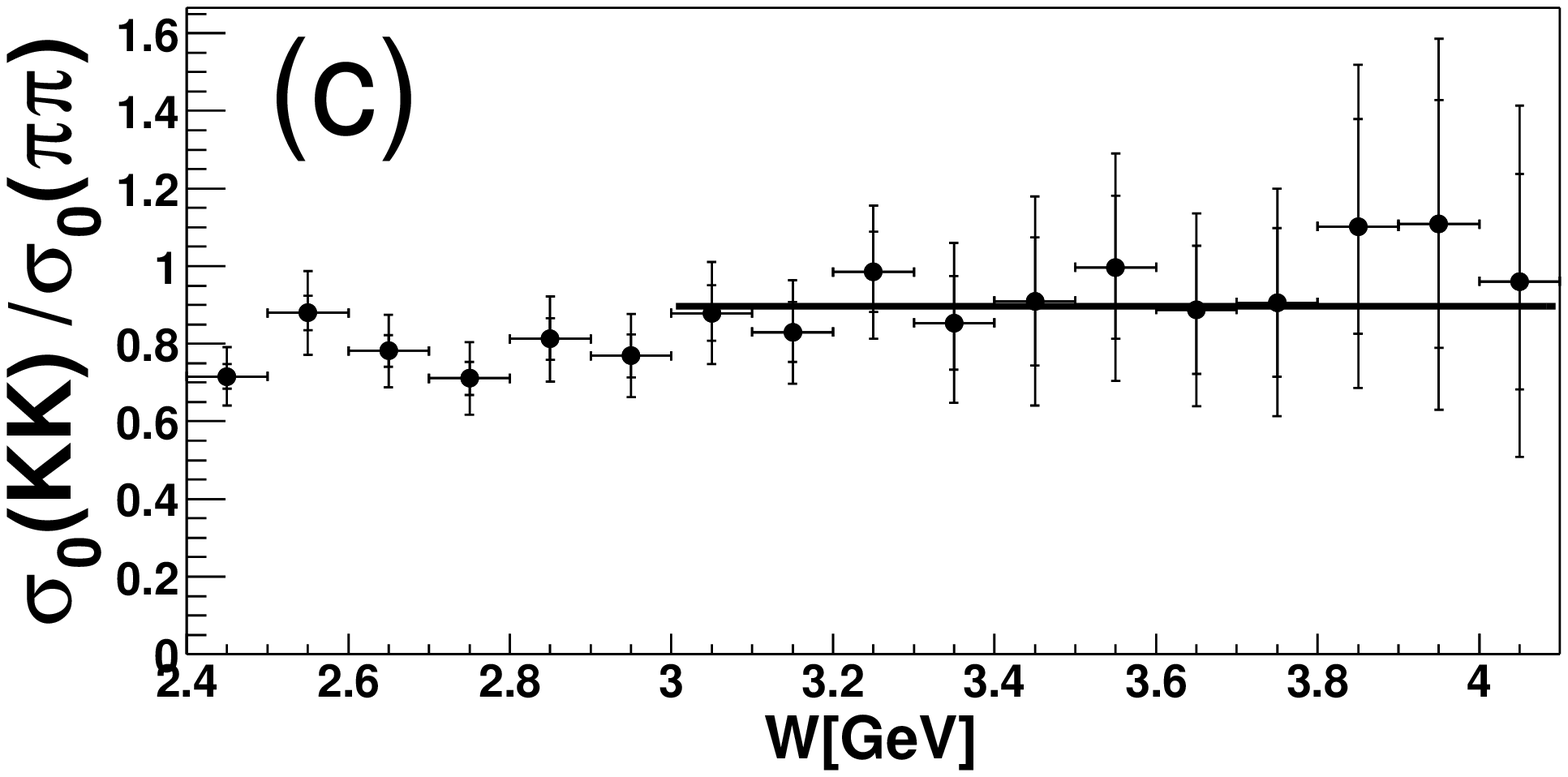,width=7cm}
  \caption{
  Cross section for
  (a) $\gamma\gamma\!\!\rightarrow\!\!\pi^{+}\pi^{-}$,
  (b) $\gamma\gamma\!\!\rightarrow\!\! K^{+}K^{-}$
  in the c.m.\ angular region $\cosre$ 
  together with a $W^{-6}$ dependence line derived from
  the fit of $s|R_M|$.
  (c) shows the cross section ratio.
  The solid line is the result of the fit
  for the data above 3$\GeV$.
  The errors indicated by short ticks are statistical only.
  }
  \label{fg_KKPIsigmaw}
\end{figure}
 \end{center}

The efficiency-corrected measured %and expected 
differential cross sections for
   $\ggpipi$ and $\ggKK$, normalized to
   the partial cross section $\sigma_0$ for $\cosre$, are shown
   in Fig.~\ref{fg:sin4} for each 100 MeV wide $W$ bin.  The partial cross sections
   $\sigma_0$ for both processes, integrated over the above scattering angle
   range, are shown in Fig.~\ref{fg_KKPIsigmaw} 
   (along with their ratio) and itemized in
   Table~\ref{table_sigma}.

%figKPI
%\twocolumn[
%]

\begin{table}[htbp]
 \caption{
   Cross sections and errors
 for the $\ggpipi$ and $\ggKK$ processes,
 in the angular range $\cosre$.}
 \vspace{0.8cm}
\begin{center}
 \begin{tabular}{c|ccc|ccc}
 \hline
 & \multicolumn{3}{c|}{$\ggpipi$} 
 & \multicolumn{3}{c}{$\ggKK$} \\
 \hline
 $W$
 & $\sigma_0$ & stat.err & syst.err
 & $\sigma_0$ & stat.err & syst.err \\
    GeV
     & nb & nb & nb
     & nb & nb & nb \\ \hline
    2.4-2.5 & 0.832 & 0.026 & 0.083  & 0.595 & 0.019 & 0.062\\
    2.5-2.6 & 0.625 & 0.024 & 0.070  & 0.549 & 0.018 & 0.059\\
    2.6-2.7 & 0.636 & 0.025 & 0.067  & 0.497 & 0.018 & 0.054\\
    2.7-2.8 & 0.530 & 0.023 & 0.059  & 0.377 & 0.016 & 0.043\\
    2.8-2.9 & 0.448 & 0.022 & 0.048  & 0.364 & 0.016 & 0.043\\
    2.9-3.0 & 0.407 & 0.021 & 0.042  & 0.313 & 0.015 & 0.038\\
    3.0-3.1 & 0.302 & 0.019 & 0.032  & 0.266 & 0.014 & 0.034\\
    3.1-3.2 & 0.247 & 0.017 & 0.026  & 0.205 & 0.013 & 0.027\\
    3.2-3.3 & 0.202 & 0.016 & 0.022  & 0.199 & 0.013 & 0.027\\
    3.3-3.4 & 0.153 & 0.016 & 0.023  & 0.130 & 0.012 & 0.022\\
    3.4-3.5 & 0.103 & 0.014 & 0.018  & 0.093 & 0.011 & 0.018\\
    3.5-3.6 & 0.093 & 0.014 & 0.017  & 0.093 & 0.011 & 0.016\\
    3.6-3.7 & 0.079 & 0.011 & 0.013  & 0.070 & 0.009 & 0.012\\
    3.7-3.8 & 0.068 & 0.011 & 0.014  & 0.062 & 0.008 & 0.011\\
    3.8-3.9 & 0.043 & 0.009 & 0.010  & 0.048 & 0.007 & 0.009\\
    3.9-4.0 & 0.035 & 0.008 & 0.010  & 0.039 & 0.007 & 0.008\\
    4.0-4.1 & 0.041 & 0.009 & 0.013  & 0.039 & 0.007 & 0.008\\
    \hline
 \end{tabular}
\end{center}
 \label{table_sigma}
\end{table}

\section{Systematic errors}
\begin{table}[h]
  \caption{
    Contributions to the systematic errors. 
    A range is shown when the error has a $W$
 dependence.}
\vspace{0.5cm}
\begin{center}
\begin{tabular}[t]{ccc}
\hline
 Source &$\pipi$ & $\KK$\\ \hline\hline

 Tracking efficiency& 4\%&4\%\\
 Trigger efficiency& 4\%&4\% \\ 
 $K/\pi$ separation&0--1\% &2--4\% \\ 
 $\mu\mu$ background subtraction&5--17\% & 0\%\\
 Non-exclusive background subtraction &4--27\% &7--20\% \\
 Luminosity function&5\% &5\% \\
 Integrated luminosity&1\% &1\% \\ 
 %\hline\hline
 %subtotal                  & 10-32\% & 10-21\% \\ \hline
 $\chi_c$ subtraction ($3.3\GeV<W<3.6\GeV$) &8--11\% &8--11\% \\ \hline
 Total&10--33\% &10--21\% \\ \hline
\end{tabular}
\end{center}
\label{table:syserr}
\end{table}

\indent
The dominant systematic errors are listed in Table~\ref{table:syserr}.  
   The uncertainty due to trigger efficiency
   is estimated by comparing the yields of $\ggmumu$ 
   in real and simulated data~\cite{aafhb} after accounting for 
   the background from $\ee\to\mumu\, n\gamma$ events
   (varying with $W$ from 0.5--4.6\%), 
   which have the same topology~\cite{mumukk}.
   The uncertainty in the relative muon
   identification efficiency between real and simulated data is used to
   determine the error associated with the residual $\mumu$ subtraction
   from the $\pipi$ sample.  
We use an error of
   100\% of the subtracted value for the non-exclusive background subtraction.
   We allow the number of $\chi_{cJ}$ events to fluctuate by up to 20\% of the
   measured excess to estimate the error due to the $\chi_c$ subtraction 
   that is applied for the energy bins in 
   the range $3.3 \GeV < W < 3.6 \GeV$.  
   The total $W$-dependent systematic error is
   10--33\% (10--21\%) for the $\ggpipi$ ($\ggKK$) cross section.

\section{Discussion}

For $\ggpipi$ and $\ggKK$, Brodsky and Lepage
   predicted a $\sinnf$ dependence of the differential 
   cross section~\cite{Lepage} as seen in Eq.~\ref{eq:brod},
   and Brodsky and Farrar (BF) predicted a $W^{-6}$ dependence of the cross
   section~\cite{scaling}.  
   The soft hadron exchange (``handbag'') model by DKV~\cite{handbag} 
   predicts an expression for the differential cross section similar to Eq.~\ref{eq:brod}
\begin{eqnarray}
 \hspace{2cm}\frac{d\sigma}{d \costh}(\ggMM) \approx
	\frac{8\pi\alpha^{2}}{s}\frac{|R_{2M}(s)|^{2}}{\sinf},
\label{eq:dt}
\end{eqnarray}
\noindent where $M$ denotes either a pion or a kaon.
   Here, they introduce the ``annihilation form factor''
   $R_{2M}(s)$, which can be determined experimentally.

   Fig.~\ref{fg:sin4} shows normalized angular distributions
   as a function of $W$, for $\pipi$ and $\KK$ modes.
   The solid curves indicate the expectations from a $\sinnf$ 
   behavior predicted by BL and DKV models
   \begin{eqnarray}
     \hspace{3.5cm}
     \frac{1}{\sigma_0}
     \frac{d\sigma}{d\costh}
     %= 
     %\frac{\sinnf}{\int^{0.6}_{0}\sinnf d\costh}
     =
     \frac{1.227}{\sinf}.
     \label{eq:sin4}
   \end{eqnarray}
   These curves match the data well
   for $\highr$, but are more shallow than the
   data for $W < 3.0\,{\rm GeV}$.  This disagreement may arise from the
   presence of one or more lower-energy resonances:  for example, Belle has
   reported a resonance at $(2327 \pm 6 \pm 6)\,{\rm MeV}/c^2$
   with a width of $(275 \pm 12 \pm 20)\,{\rm MeV}/c^2$ in a study of
   $\ggKK$ with $W < 2.4\,{\rm GeV}$~\cite{uehara}.
   Small momentum transfer from the photons to the pions, $-t < 3\GeV^2$ 
   at smaller scattering angles, may also cause this disagreement
   for $W < 3.0\,\GeV$
   through some unknown non-perturbative or hadronic effects.

We extract the annihilation form factor $R_{2M}(s)$ directly from the data,
   by integrating the observed cross sections over $\cosre$ 
   in the range of $\highr$,
   where Eq.~\ref{eq:sin4} fits the data quite well.  
   Since the observed values of $s|R_{2M}(s)|$ are almost $W$-independent,
   we obtain
   $s|R_{2\pi}| = \ValsRPI$ and $s|R_{2K}| = \ValsRKK$ 
   %for $\ggpipi$ and $\ggKK$, respectively,
   by fits to the data in this $W$ range.
   Fig.~\ref{fg_KKPIsigmaw} shows the observed cross sections
   in comparison with the recent ALEPH data~\cite{aleph}.
   The lines indicate expectations with the 
   $s|R_{2M}(s)|$ values from the fit above.
   Our data are consistent with $W^{-6}$ behavior
   predicted by BL~\cite{Lepage} and BF~\cite{scaling} models.
   
   We can also directly obtain the power $n$ of the $W$-dependence
   ($\sigma_0\propto W^n$) from the data.
   We find $n=\ValPIPowers$ for $\pipi$ and $\ValKKPowers$ for $\KK$,
   for $\highr$.
   The first error is statistical, and the second is systematic.
We conservatively estimate the systematic error of $n$ by
artificially deforming the measured cross section values under the
assumption that systematic errors are strongly point-to-point correlated:
we shift the $\sigma_0$ values at the two end bins by $\pm 1.5$ and
$\mp 1.5$ times the systematic error, respectively, whereas each intermediate point
is correspondingly moved so that its shift follows a linear function of $W$
times its systematic error. The average of the observed deviations of
$n$ from its original value is taken as a final systematic error.
   The results show a hint of somewhat steeper dependence than $W^{-6}$.

Fig.~\ref{fg_KKPIsigmaw}(c) shows the cross section ratio 
$\sigma_0(\ggKK)/\sigma_0(\ggpipi)$
as a function of $W$.
The ratio is energy independent 
for $\highr$,
in accordance with the QCD prediction. 
The obtained value of the ratio is
$\ValSigmaRatio$. It is consistent with the value of 1.08 predicted by
Benayoun and Chernyak (BC)~\cite{chernyak} and 
significantly below 2.23 following from the BL calculation~\cite{Lepage}
and using the current values of the kaon and pion decay constants~\cite{PDG}.
The value predicted by BC is a consequence of consistent consideration of the
SU(3) breaking effects using the different wave functions for pions and kaons
derived from the QCD sum rules. This compensates for the partial account of 
the SU(3) breaking by BL who used the same wave functions for pions and kaons
so that the cross section ratio is equal to the fourth power of the ratio
of the kaon and pion decay constants.

\section{
TWO-PHOTON DECAY WIDTH OF $\chi_{cJ}$ RESONANCES
\label{sec:chic}}

\begin{table*}[htbp]
\caption{ 
  Results for the product of the two-photon
  decay width and the branching fraction,
  $\Ggg(\chi_{cJ})\mathcal{B}(\chi_{cJ}\to\MM)$.
  The second column gives the observed $\chi_{cJ}$ yields
  in the $W$ region of 3.34--3.44 GeV (3.54--3.58 GeV) 
  for $\chi_{\!c0}$ ($\chi_{c2}$).
  The first and second errors for $\Ggg\mathcal{B}$
  are statistical and systematic, respectively.
  \vspace{0.5cm}
}
\begin{tabular}{cccc}
  %\begin{tabular*}{0.8\textwidth}%
  %{@{\extracolsep{\fill}}cccc}
\hline
%& Number of events & $\Ggg(\chi_{cJ})\mathcal{B}(\chi_{cJ}\!\to\!\MM)$[eV] & significance \\
%\parbox{4cm}{\hfill}& \parbox{5cm}{Number of events} & $\Ggg(\chi_{cJ})\mathcal{B}(\chi_{cJ}\!\to\!\MM)$[eV] & 
%\parbox{4cm}{Significance} \\
& Number of events & $\Ggg(\chi_{cJ})\mathcal{B}(\chi_{cJ}\!\to\!\MM)$[eV] & 
Significance \\
\hline
$\gg\to\chi_{c0}\to\pipi$ & $\ValPINCHIAwithE$ & $\ValPIGggBrA$ & $\ValPISIGA$ \\
$\gg\to\chi_{c0}\to\KK$ & $\ValKKNCHIAwithE$ & $\ValKKGggBrA$ & $\ValKKSIGA$ \\
$\gg\to\chi_{c2}\to\pipi$ & $\ValPINCHIBwithE$ & $\ValPIGggBrB$ &$\ValPISIGB$ \\
$\gg\to\chi_{c2}\to\KK$ & $\ValKKNCHIBwithE$ & $\ValKKGggBrB$ &$\ValKKSIGB$ \\
%\omit \hskip 0.25\textwidth & \hskip 0.25\textwidth & \hskip 0.25\textwidth & \hskip 0.25\textwidth \\
\hline
\end{tabular}
\label{tb:GggBr}
\end{table*}

The measured yields of $\chi_{c0}$ and $\chi_{c2}$ events
can be used
to extract
the two-photon decay width of each resonance
using the formula
\begin{eqnarray}
  \hspace{1cm}
\Ggg(\chi_{cJ}) = 
\frac
{YM_{cJ}^{2}}
{
4(2J\!+\!1)\pi^2
\Lgg(M_{cJ})
\epsilon
\mathcal{B}(\chi_{cJ}\!\to\!\MM)\Ldt
}
\end{eqnarray}
\noindent
   where $Y$ and $M_{cJ}$ are the yield and mass, respectively, of the 
   corresponding spin-$J$ $\chi_{cJ}$
   resonance (where the mass is taken from Ref.~\cite{PDG}).
   $\epsilon$ denotes the product of the
   detection efficiency and acceptance for the resonance decays.
   The extracted values of $\Ggg(\chi_{cJ})
   \mathcal{B}(\chi_{cJ} \to M^+M^-)$ are summarized in Table~\ref{tb:GggBr}.  
   The ratios $\Ggg\mathcal{B}(\KK)/\Ggg\mathcal{B}(\pipi)$ are consistent 
   with the known $\mathcal{B}(\KK)/\mathcal{B}(\pipi)$~\cite{PDG}
   ratios for both resonances.
   Therefore we combine the measurements of the widths from the two modes 
   to obtain 
$\Ggg(\chi_{c0})=\ValCombGggA$
and
$\Ggg(\chi_{c2})=\ValCombGggB$
where the error associated with the branching fraction
(taken from ~\cite{PDG}) is calculated
assuming independent uncertainties for the $\pi^+\pi^-$ and $K^+K^-$ modes.
These values are consistent with the two-photon decay widths of
$\chi_{c0}$ and $\chi_{c2}$ following from the total widths and
branching ratios into two photons obtained in 
previous experiments~\cite{PDG}.

\section{Conclusion}
Using $\ValIntLumi$ of data collected with the Belle
detector at KEKB,
we have measured with high precision
the cross sections for the
$\ggpipi$
and
$\ggKK$
processes
in the two-photon c.m.\ energy range
$\massr$ and angular range of $\cosre$.

   The angular differential cross sections 
   for those processes
   show a $\sinnf$ behavior in the range of $\highr$,
   which is predicted by QCD.
   The energy dependence of the differential cross section is
   also consistent with the QCD prediction of $\sigma\propto W^{-6}$.
   %The extracted $s$-scaled annihilation form factors 
   %in this range are
   %$s|R_{2\pi}| = \ValsRPI$ and $s|R_{2K}| = \ValsRKK$.
   The partial cross section ratio $\sigma_0(\ggKK)/\sigma_0(\ggpipi)$,
   in $\cosre$
   is measured to be $\ValSigmaRatio$ for $\highr$
   in accordance with the prediction based on the QCD sum rules and
   SU(3) symmetry breaking.

We have observed the $\chi_{c0}$ and $\chi_{c2}$ resonances decaying
to $\pipi$ and $\KK$ final states and measured the products of
their two-photon decay widths and two-hadron branching fractions for
the first time.

\newpage
\textbf{Acknowledgements}\\
% Please paste this acknowledgement into your latex file. 
% most recent update: 04-07-15
%***** Acknowledgments *****
% use these two starting with pub # 98
%----------- Long version, for most papers ----------- 
We thank the KEKB group for the excellent operation of the
accelerator, the KEK Cryogenics group for the efficient
operation of the solenoid, and the KEK computer group and
the National Institute of Informatics for valuable computing
and Super-SINET network support. 
We are grateful to V. Chernyak for fruitful discussions.
We acknowledge support from
the Ministry of Education, Culture, Sports, Science, and
Technology of Japan and the Japan Society for the Promotion
of Science; the Australian Research Council and the
Australian Department of Education, Science and Training;
the National Science Foundation of China under contract
No.~10175071; the Department of Science and Technology of
India; the BK21 program of the Ministry of Education of
Korea and the CHEP SRC program of the Korea Science and
Engineering Foundation; the Polish State Committee for
Scientific Research under contract No.~2P03B 01324; the
Ministry of Science and Technology of the Russian
Federation; the Ministry of Education, Science and Sport of
the Republic of Slovenia;  the Swiss National Science Foundation; the National Science Council and
the Ministry of Education of Taiwan; and the U.S.\
Department of Energy.

\end{document}